\documentclass[letterpaper]{article} 
\usepackage{aaai2026}  
\usepackage{times}  
\usepackage{helvet}  
\usepackage{courier}  
\usepackage[hyphens]{url}  
\usepackage{graphicx} 
\pdfoutput=1
\usepackage{natbib}  
\usepackage{caption} 
\usepackage{algorithm}
\usepackage{algorithmic}

\usepackage{newfloat}
\usepackage{listings}
\DeclareCaptionStyle{ruled}{labelfont=normalfont,labelsep=colon,strut=off} 
\floatstyle{ruled}
\newfloat{listing}{tb}{lst}{}
\floatname{listing}{Listing}

\usepackage[utf8]{inputenc} 
\usepackage[T1]{fontenc}    
\usepackage{booktabs}       
\usepackage{amsfonts}       
\usepackage{nicefrac}       
\usepackage{microtype}      
\usepackage{xcolor}         
\usepackage{tcolorbox}
\usepackage{xspace}
\usepackage{enumitem}
\usepackage{amsthm}
\theoremstyle{acmdefinition}

\usepackage{multirow}
\usepackage{tabularx}
\newcolumntype{C}[1]{>{\centering\arraybackslash}p{#1}}
\usepackage{subcaption}
\usepackage{xfrac}

\newcommand{\todo}[1]{\textcolor{black}{#1}}
\newcommand{\tool}{\textsc{SynPrune}\xspace}

\newcommand{\ie}{\textit{i.e.,}\xspace}
\newcommand{\eg}{\textit{e.g.,}\xspace}

\title{Uncovering Pretraining Code in LLMs: A Syntax-Aware Attribution Approach}

\author{
    Yuanheng Li\textsuperscript{\rm 1}\thanks{Both authors contributed equally to this work.},
    Zhuoyang Chen\textsuperscript{\rm 1}\footnotemark[1],
    Xiaoyun Liu\textsuperscript{\rm 1},
    Yuhao Wang\textsuperscript{\rm 1},\\
    Mingwei Liu\textsuperscript{\rm 2},
    Yang Shi\textsuperscript{\rm 1},
    Kaifeng Huang\textsuperscript{\rm 1}\thanks{Kaifeng Huang is the corresponding author.},
    Shengjie Zhao\textsuperscript{\rm 1}
}

\affiliations{
    \textsuperscript{\rm 1}Tongji University\\
    \textsuperscript{\rm 2}Sun Yat-sen University
}

\usepackage{bibentry}

\begin{document}

\maketitle




\begin{abstract}

As large language models (LLMs) become increasingly capable, concerns over the unauthorized use of copyrighted and licensed content in their training data have grown, especially in the context of code. Open-source code, often protected by open-source licenses (e.g., GPL), poses legal and ethical challenges when used in pretraining. Detecting whether specific code samples were included in LLM training data is thus critical for transparency, accountability, and copyright compliance.

We propose \tool, a syntax-pruned membership inference attack method tailored for code. Unlike prior MIA approaches that treat code as plain text, \tool leverages the structured and rule-governed nature of programming languages. Specifically, it identifies and excludes consequent tokens that are syntactically required and not reflective of authorship from attribution when computing membership scores. Experimental results show that \tool consistently outperforms the state-of-the-arts.
Our method is also robust across varying function lengths and syntax categories. 

\end{abstract}



\section{Introduction}

The fundamentals of large language models (LLMs) stems from the vast pre-training datasets. However, the copyright issue in the training corpora deserves serious attention. 
Recently, several legal disputes have arisen concerning the use of copyrighted materials in training data~\cite{openaisue, newssue}. 
Respecting copyright and intellectual property is essential, not only as a legal obligation enforced by governments, but also as a means of fostering supportive and respectful environments for content creators. As LLMs proliferate and AI-generated content grows, safeguarding human creativity is vital to preserving unique and diverse perspectives.



LLMs' coding abilities have enabled applications from autonomous agents to automated coding. 
However, these coding abilities stem from large-scale training on open-source code, which often comes with specific usage obligations. However, such code is typically governed by specific usage obligations.
A well-known example is the copyleft license (\eg~GPL~\cite{gnu}). 
The GPL was introduced in 1989 as part of the GNU Project~\cite{gnu}. Eighteen years later, we witnessed the first recognized lawsuit for a GPL violation~\cite{lvcriminaldefense, Informationweek}, when Monsoon Multimedia was sued by the Software Freedom Conservancy.
The following decades have seen many GPL conflicts and litigations~\cite{gpllitigation}, which underscores the importance of respecting copyrighted content~\cite{xu2024llms}. 
We anticipate that similar open-source litigation may soon target LLM vendors.  
In fact, disputes over the use of code for pretraining have already emerged~\cite{githubsue}, and few have reached resolution due to the protracted nature of litigation. As a result, the transparency and accountability questions in LLMs remain unanswered.

Several studies attempt to answer the transparency and accountability issues in LLMs. Zhou et al.~\cite{zhou2024dpdllm} analyzed books, websites, and Wikipedia to assess potential copyright violations. Recent benchmarks~\cite{chen2024copybench, liu2024shield} support evaluation of both literal copying (\eg~verbatim reproduction) and non-literal copying (\eg~replicated plots or characters) from LLM outputs. There are also proactive measures known as copyright traps~\cite{shilov2024mosaic, meeus2024copyright}, which embed crafted copyrighted content to enable precise copyright violation detection. However, this approach is limited by the use of pre-inserted traps rather than those derived from real-world scenarios.


Another approach involves applying membership inference attacks (MIAs) to LLMs, which initially determine whether a specific data point was part of the training dataset of a machine learning model
~\cite{shokri2017membership, song2021systematic, song2019privacy, huang2024general, yeom2018privacy, leino2020stolen, long2018understanding, nasr2019comprehensive, sablayrolles2019white, salem2018ml, truex2019demystifying}, thereby exposing potential privacy risks. 
MIAs in LLMs share the same intuitions of MIAs in machine learning that models memorize data (overfitting)~\cite{song2019auditing, carlini2021extracting, tirumala2022memorization}. Tirumala et al.~\cite{tirumala2022memorization} find that models memorize nouns and numbers, and that larger models can memorize a larger portion of the data before over-fitting. 
These methods assume access to pretraining data to train shadow or reference models. Yang et al.~\cite{yang2024gotcha} presented \textsc{GotCha}, a membership inference attack using surrogate models tailored for code LLMs.
However, training such models is computationally expensive, and pretraining data is often unavailable, especially for commercial LLMs.
Differently, several MIAs on LLMs typically exploit the model's token-level outputs and probabilities~\cite{carlini2021extracting}, often in combination with techniques such as synonym substitution to leverage neighboring data~\cite{mattern2023membership}, detection of low-probability tokens~\cite{shi2023detecting}, or calibration of token probabilities using term frequency divergence~\cite{zhang2024pretraining}. 
However, their effectiveness is still limited, as Duan~\cite{duan2024membership} found that MIAs on LLMs perform only marginally better than random guessing. In addition, most existing MIAs for LLMs target general natural language text rather than code.

Although existing text-based MIA techniques can be applied to code, their full potential remains untapped. A key distinction between code and natural language is the presence of formal syntax and structured constructs. When developers write code, they reason about task logic while adhering to the programming language's syntax rules. We hypothesize that \textit{the effectiveness of MIA techniques on code-related LLMs is limited by their failure to leverage these syntactic characteristics}. Through our observations, we find that certain syntax elements (\eg~Data models, expressions, and statements) follow syntax conventions, rather than expressing individual authorship. That is, once specific syntax \textit{conditions} are met, certain \textit{consequent tokens} must or are highly likely to appear. Therefore, \textit{consequent tokens} should be excluded from attribution when computing MIA scores. Based on this insight, we propose \tool, a syntax-pruned membership inference attack (MIA) method tailored for code. We summarize a set of \todo{47} syntax conventions derived from the official Python language reference.

We evaluate \tool on a curated benchmark of Python functions that are authentic and verified, as existing benchmark on code only provides synthetic or assumed members and non-members. In our benchmark, Member functions are sourced from the Pile dataset, which is widely used in the pretraining of many LLMs. Non-member functions are collected from sources published after the release cut-off dates of the evaluated LLMs, ensuring that none of the models have encountered them during training. 
Experimental results show that \tool outperforms state-of-the-art MIA techniques, achieving an average AUROC improvement of \todo{15.4\%} across \todo{four} models and three member-non-member ratios. Additionally, we demonstrate that \tool remains robust across varying function lengths. Through ablation studies, we assess the contribution of each syntax convention category. 


\textbf{Contribution.} We make the following contributions in detecting pretraining code in LLMs. 

\begin{itemize}
    \item We propose \tool, a syntax-pruned membership inference attack method for code, which excludes from attribution tokens that are inherently determined by Python syntax conventions.
    \item We present a benchmark comprising authentic and verifiable member and non-member functions, offering a reliable ground truth often absent in existing approaches.
    \item We evaluate the effectiveness of \tool against state-of-the-art approaches, examine its robustness across varying function lengths, and analyze the impact of different syntax convention categories through ablation studies.
\end{itemize}






\section{Related work}

\textbf{MIA in Machine Learning.} Membership inference attacks (MIAs) have gained traction as tools to audit privacy risks in machine learning~\cite{song2021systematic, song2019privacy, huang2024general, yeom2018privacy, leino2020stolen, long2018understanding, nasr2019comprehensive, sablayrolles2019white, salem2018ml, truex2019demystifying}. Shokri et al.~\cite{shokri2017membership} introduced a shadow model-based approach assuming knowledge of the model architecture and data distribution. Subsequent works relaxed or challenged these assumptions. Salem et al.~\cite{salem2018ml} used diverse shadow model architectures; Liu et al.~\cite{liu2022membership} correlated loss trajectories between distilled models and target model; Choquette et al.~\cite{choquette2021label} assumed confidence scores are unavailable. Rezaei et al.~\cite{rezaei2021difficulty} highlighted evaluation pitfalls between overfitted and well-trained models. Kazmi et al.~\cite{kazmi2024panoramia} assume synthetic non-member data to be consistent with queried non-member data. Ye et al.~\cite{ye2022enhanced} formalized attack assumptions in a unified framework. Besides, the shadow model framework has seen other advances. Yeom et al.~\cite{yeom2018privacy} showed that overfitting alone enables effective attacks, where a simple threshold on prediction confidence can match the accuracy of complex attack models. Song et al.~{song2021systematic} highlighted the lack of per-sample analysis, prompting difficulty calibration methods that adjust membership scores based on sample hardness~\cite{watson2021importance, he2024difficulty}. Carlini et al.~\cite{carlini2022membership} emphasized evaluating attacks at low FPRs rather than relying on average-case metrics.

\textbf{MIA in Language Models.} Song et al.~\cite{song2019auditing} investigated how deep learning-based text generation models memorize training data, while Carlini et al.~\cite{carlini2021extracting} demonstrated that large language models can leak verbatim training sentences. While literal (verbatim) memorization is easier to detect, identifying non-literal memorization remains challenging. Most approaches focus on modeling the likelihood of word sequences in context. Duan~\cite{duan2024membership} found that MIAs perform only slightly better than random guessing in LLMs, due to the inherently fuzzy boundary between members and non-members.
To address this, Mattern et al.~\cite{mattern2023membership} proposed generating neighboring data and comparing their distribution to infer membership. Zhang et al.~\cite{zhang2024pretraining} calibrated token probabilities by measuring the divergence between within-document and corpus-level term frequencies, hypothesizing that higher divergence signals greater information content. Shi et al.~\cite{shi2023detecting} introduced a reference-free method, MIN-K\% prob, which assumes members are less likely to contain low-probability tokens compared to non-members.

\begin{table*}[!t]
    \centering
    \footnotesize
    \caption{List of the Collected Syntax Conventions (\texttt{[SP]}, \texttt{BR}, and \texttt{[IND]} denote the space, line break, and indentation, resp.)}\label{table:syntax}
    \begin{tabular}{m{2cm}m{5cm}m{1.5cm}m{4cm}m{3cm}}
        \toprule
        \textbf{Category} &\textbf{Syntax Nodes}  & \textbf{Example Syntax Node} & \textbf{Conditional Token} & \textbf{Consequent Token} \\
        \midrule
        
        \multirow{3}{*}{Data Model}  & List, Slice, Dict, Set, String, Bytes, Object, Tuple 
        & List & `[' & `]' \\
        & & Dict &  `\{' & `\}' \\
        & & Tuple & `(' & `)' \\\hline

        \multirow{4}{*}{Expression} 
        
        & \multirow{4}{*}{\shortstack[l]{call, lambda, conditional, comprehension,\\ Chained Comparison}} 
        
        & \multirow{2}{*}{call} & \textit{identifier}( & `)' \\
         &  &      & \textit{identifier}(\textit{identifier} & `,' \\ \cmidrule{4-5}
        
        &  & conditional & \textit{expr} \texttt{if} \textit{cond} & \texttt{else} \\
        & & \shortstack[l]{chained\\comparison} & \textit{expr comp\_op expr} & \textit{comp\_op} \\\hline

        \multirow{2}{*}{Single Stmts}  &  \multirow{2}{*}{\texttt{import}, \texttt{assert}, \texttt{global}} & \multirow{2}{*}{\texttt{import}} & \texttt{import} \textit{module}  & \texttt{as} \\

        & & & \texttt{from} \textit{module} & \texttt{import} \\
        \hline
         \multirow{6}{*}{Compound Stmts} & \multirow{6}{*}{\shortstack[l]{\texttt{if}, \texttt{for}, \texttt{try}, \texttt{with}, \texttt{class}, function,\\\texttt{while}, \texttt{match}}} & \multirow{2}{*}{\texttt{for}}  
         & \texttt{for} \textit{target\_list}  & \texttt{in}  \\

         &  &  & \texttt{for} \textit{target\_list} \texttt{in} \textit{starred\_list}   &  `:', \texttt{[SP]}, \texttt{[IND]},  \texttt{[BR]} \\\cmidrule{4-5}

         & & \multirow{1}{*}{\texttt{if}} & \texttt{if} \textit{assignment\_expression} & `:'\\
         & & \texttt{try} & \texttt{try:} \textit{suite} \texttt{except} \textit{expression}  &  \texttt{as} \\
         &   & function & \texttt{def} \textit{funcname}(\textit{identifier} & `,', `)' \\
         &   & \multirow{1}{*}{\texttt{with}} & \texttt{with} \textit{with\_item} & \texttt{as} \\

         &   & \multirow{1}{*}{\texttt{while}} & \texttt{while} \textit{condition} & `:'\\
         & & \multirow{1}{*}{\texttt{match}} & \texttt{match} \textit{subject} & `:' \\
        \bottomrule
    \end{tabular}
\end{table*}

\textbf{Copyright Sources in Language Models.} Xu et al.~\cite{xu2024llms} found that LLMs fail to respect copyright embedded in user inputs. Zhou et al.~\cite{zhou2024dpdllm} analyzed copyrighted sources such as books, websites, and Wikipedia to assess potential copyright violations. Recent benchmarks~\cite{chen2024copybench, liu2024shield} support evaluation of both literal copying (\eg~verbatim reproduction) and non-literal copying (\eg~replicated plots) in LLM outputs. Document memorization in LLMs is another challenge. Meeus et al.~\cite{meeus2024copyright} observed that LLMs do not memorize enough to enable reliable document-level membership inference and proposed document-specific copyright traps to support such inference. 
Shilov et al.~\cite{shilov2024mosaic} introduced fuzzy copyright traps designed to evade deduplication and persist through training. Interestingly, despite the well-established nature of code copyright—supported by numerous litigations and enforcement cases, there has been relatively little research on copyright issues in code-generating LLMs.
Yang et al.~\cite{yang2024gotcha} presented \textsc{GotCha}, a membership inference attack tailored for code models. Similarly, Wan et al.~\cite{wan2024does} proposed a membership inference attack targeting code completion models.








\section{Methodology}\label{sec:methodology}

We discuss the motivation for our method design as the preliminary. Then we elaborate our method \tool.

\subsection{Preliminary}






\begin{figure*}[!t]
    \centering
    \includegraphics[width=0.98\textwidth]{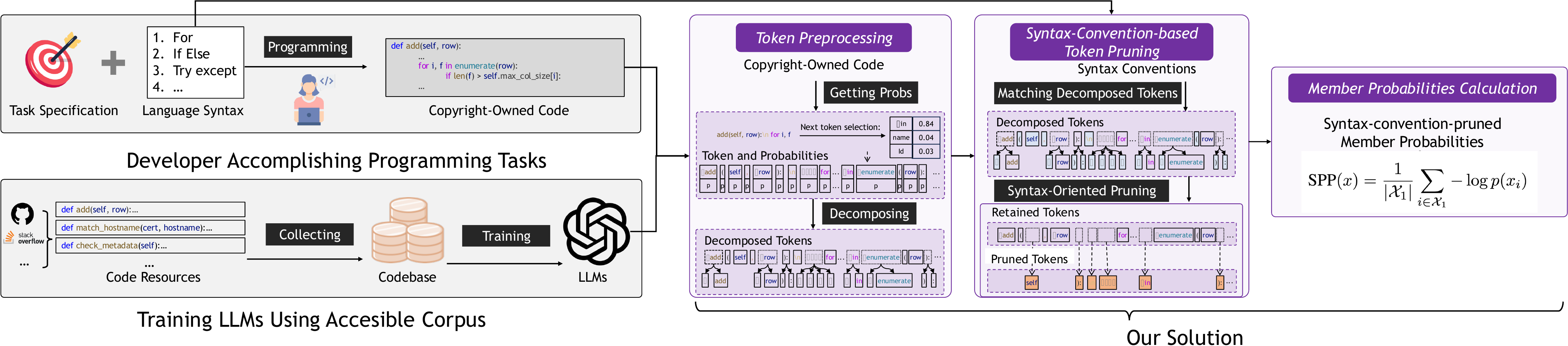}
    \caption{Illustration of our Approach Motivation and Proposed Solution}
    \label{fig:overview}
\end{figure*}

Source code is considered a ``literary work'' under copyright law. The ``literary work'' can be viewed as a sequence of tokens arranged to reflect the author's intentions. Existing MIA techniques~\cite{meeus2024copyright, yang2024gotcha, wan2024does} treat all source code tokens equally and compute membership probabilities of an LLM based on a sample of these tokens. Generally, it is recognized that elements such as variable names, data and control flows, and API usage, together with the involved variables, embody the author's intent and serve as indicators of source code authorship. However, we observe that some tokens primarily serve a grammatical purpose, functioning to complete the syntax constraint of the programming language.


\subsection{Syntax Conventions}

We chose Python as our target language for studying syntax conventions, as it is a widely used programming language with broad applications in data analysis, artificial intelligence, and scientific computing. Syntax conventions refer to the specific rules defined by the Python language that programmers \textit{must} follow. Failure to do so will result in syntax errors and prevent the code from executing. Therefore, we consider tokens that follow syntax conventions to be highly predictable patterns and exclude them from the MIA calculation.

In the example provided in our overview, the programmer completes the task by declaring a method named \texttt{add} within an existing class, following both the task specification and the language's syntax requirements. The method contains various syntax elements, such as a method declaration and a \textit{for} statement. The token \texttt{self} represents the class instance and is mandatory. Likewise, tokens such as the closing parenthesis \texttt{)} and colon \texttt{:} after the \texttt{row} parameter, along with the \texttt{[BR]} and \texttt{[IND]}, are also required by Python's grammar and must be present.

To compile a comprehensive collection of syntax conventions, two of our authors 
manually reviewed the official Python documentation~\cite{python11}. Specifically, we referred to the Python Language Reference section of the Python 3.11 documentation, released in October 2022. The authors reviewed sections of the documentation that define syntax elements, including the data model (Section 3), the import system (Section 5), expressions (Section 6), as well as statements, and function and class definitions (Sections 7 and 8). A third author is involved to resolve any arguments or disagreements.

Table \ref{table:syntax} lists our summarized syntax conventions. We categorize them into four categories according to their syntax element types. i.e., data model, expressions, single statements and compound statements. The syntax convention is defined as a tuple $\langle \text{condition}, \text{consequence} \rangle$, where $\text{condition}$ represents a prefix under which certain code elements conform to specific language syntax, and $\text{consequence}$ denotes a suffix that must or high-likely to occur given the $\text{condition}$. In total, we obtained \todo{47} syntax conventions, denoted as $\mathcal{C}$.


\begin{itemize}
    \item \textit{Data Model:} Must terminate with appropriate delimiters or accessors for syntactic completeness (\eg~closing brackets `]' or dot notation)
    \item \textit{Expressions:} Must follow function identifiers and include a parenthesized, comma-separated list of arguments (\eg~`)').
    \item \textit{Simple Statement}: Probably with aliases when including external packages (\eg~\texttt{import} \textit{module} \texttt{as})
    \item \textit{Compound Statement:} These statements must be followed by a colon (`:') and an indented block. The enclosed statements start with an indentation. The compound statements are probably extended with additional branches such as \texttt{elif}, \texttt{else}, \texttt{except}, \texttt{except*}, or \texttt{finally}. The \texttt{with} statement is typically used with aliases when wrapping code defined by a context manager. Classes and methods are defined using parameters enclosed in parentheses, followed by a colon. Method definitions begin with a \texttt{self} parameter to refer to the instance.
\end{itemize}

\subsection{Syntax-Pruned MIA}

We propose \tool, a syntax-pruned membership inference attack method for inferencing code members. We hypothesize that while programmers express creativity in writing code, certain tokens inevitably occur due to syntax conventions, rather than as a reflection of individual authorship. Our approach can be divided into three phases. 

The first phase is \textit{token preprocessing}. We begin by feeding the target querying code (\eg~a function) into the evaluated LLM. Using the model's embedded tokenizer, we obtain the tokenized sequence along with the predicted probability (derived from the \textit{logits}) for each token. The resulting tokens are denoted as $\mathcal{X}$, where each token $x_i \in \mathcal{X}$ represents the token at position $i$. These tokens are split according to the tokenizer's internal logic, such as byte pair encoding (BPE)~\cite{shibata1999byte} or similar subword segmentation algorithms~\cite{li2021char}. However, to align with our predefined syntax conventions, we further decompose certain tokens. If a token contains sub-tokens that match any elements defined as $\text{condition}$ or $\text{consequence}$ in the constraint set $\mathcal{C}$, we split it accordingly. Concretely, let $\mathcal{X} = \{ x_1, x_2, ..., x_n \}$ be the original tokens, $\mathcal{X}^{\prime} = \{ x_{ij}\}$ be the decomposed sub-tokens. We define the split function:

\begin{equation}
\mathrm{Split} : \mathcal{X} \rightharpoonup \mathrm{List}(\mathcal{X}')
\end{equation}

where each $x_i \in \text{dom}(\mathrm{Split})$ is mapped to an ordered list of sub-tokens $[x_{i1},x_{i2}, ..., x_{im}]$. \text{dom} is the domain function that includes the set of inputs for function. Each original token is equal to the string concatenation of its sub-tokens:

\begin{equation}
x_i = x_{i1}  \oplus x_{i2} \oplus ... \oplus x_{im}
\end{equation}

The complete set of sub-tokens is 
\begin{equation}
\mathcal{X}^{\prime} = \bigcup_{x_{i} \in \text{dom}(\mathrm{Split})} \mathrm{Split}(x_i)
\end{equation}

The second phase is \textit{syntax-convention-based token pruning}. Given the decomposed sub-tokens $\mathcal{X}^{\prime}$, we examine each sub-token $x_{ij}$ for tokens $x_i \in \text{dom}(\mathrm{Split})$, and check whether it matches either $c.\text{condition}$ or $c.\text{consequence}$ for each constraint $c \in \mathcal{C}$. The matching process first aligns a candidate token with $c.\text{consequence}$ and records its token offset (\ie, \texttt{lineno} for the starting character position and \texttt{col\_offset} for the token length). Once a match is found, \tool traverses the Abstract Syntax Tree (AST) parsed using Python's \texttt{ast} module to locate the corresponding AST node based on the recorded character position and token length. Finally, \tool validates the associated conditional tokens according to the AST hierarchy.

If all sub-tokens $[x_{i1}, x_{i2}, \dots, x_{im_i}]$ match elements in $\mathcal{C}$, we label the original token $x_i$ as $0$ (i.e., pruned), otherwise as $1$ (i.e., retained). The function $\ell$ assigns a binary label to each token $x_i \in \mathcal{X}$, where $\ell(x_i) \in \{0, 1\}$. 

\begin{equation}
\ell : \mathcal{X} \rightarrow \{0,1\}
\end{equation}


The third phase is \textit{member probabilities calculation}. The \underline{S}yntax-\underline{P}runed MIA \underline{P}robabilities (SPP) is defined as:

\begin{equation}
\text{SPP}(x)=\frac{1}{|\mathcal{X}_{1}|}\sum_{i \in \mathcal{X}_{1}} -\log p(x_i)  
\end{equation}

where $ \mathcal{X}_1 = \{x_i \in \mathcal{X} \mid \ell(x_i) = 1\} $. We set a threshold $\epsilon$ to determine code membership. If $\text{SPP}(x) > \epsilon$, the sample is predicted as a member. Otherwise, it is classified as a non-member.

\section{Benchmark Construction}


Existing MIA benchmarks, such as WikiMIA~\cite{shi2023detecting} and MIMIR~\cite{duan2024membership}, are designed to evaluate membership inference attacks on general textual data, rather than source code.
The source code benchmark used in Yang et al.\cite{yang2024gotcha} consists of members and non-members randomly sampled from CodeXGLUE\cite{lu2021codexglue}, with labels assigned afterward. We argue that such a design fails to reflect the real-world distinctions between members and non-members. To address this, we constructed an authentic and verifiable benchmark in which members are traceable to the public datasets declared by the evaluated models, and non-members are created after the release cut-off date of the LLMs. Table~\ref{tab:benchmark} presents the statistics of our collected benchmark on the Python language. Specifically, we collected member and non-member samples as follows.


\textbf{Members.} We leveraged the Pile~\cite{pile}, a large-scale open-source dataset composed of multiple smaller datasets, including 7.6\% of a repository dataset sourced from GitHub~\cite{thepile}. The Pile was released in 2021, serving as the training dataset for many LLMs, including Pythia~\cite{biderman2023pythia}, GPT-Neo~\cite{black2021gptneo}, StableLM~\cite{stabilityai2023stablelm}, etc. We randomly sampled 1,000 Python functions by selecting 10 functions from every 100 consecutive entries in the Pile dataset.


\textbf{Non-members.} For non-members, we searched GitHub using a customized query for Python repositories, limiting the creation time to after January 1, 2024 (All four evaluated LLMs had been released prior to this date.). We sorted the repositories by star count in descending order to ensure quality. We then extracted 10 Python functions from each repository before proceeding to the next one, continuing this process across 100 repositories to collect a total of 1,000 functions. To ensure these functions were genuinely original and not cloned from pre-existing sources, we implemented a rigorous verification process. First, we parsed each candidate function's code using Python's \texttt{ast} module to extract its name, variable names, and function calls. These elements were then used to build search queries for the GitHub API. The verification relied on three heuristics: (1) searching for the exact function name to identify direct duplicates; (2) searching by internal variable names to detect refactored code reuse; and (3) searching for the complete string of function calls to find logic similarities. Two of the authors conducted peer reviews on the search results to ensure that all 1,000 functions were original and created after January 2024.

Table~\ref{tab:benchmark} summarizes the statistics of our collected benchmark. The benchmark includes 214 non-member function files with an average of 25.34 lines of code (LOC). For member functions, file counts are unavailable as this information was not provided in the Pile dataset.

\begin{table}[!t] 
    \centering
    \footnotesize
        \caption{Benchmark Statistics}\label{tab:benchmark}
    \begin{tabular}{m{1.2cm}m{1.2cm}m{1.2cm}m{1.5cm}}
    \toprule
    \multirow{1}{*}{\textbf{Label}}   & \textbf{Files (\#)}  & \textbf{Func. (\#)}   & \textbf{Aver. LOC (\#)}   \\ 
    \midrule
    \multirow{2}{*}{Members}   & \multirow{2}{*}{-} & \multirow{2}{*}{1000} & \multirow{2}{*}{14.03}  \\ 
     & & & \\ 
    \cmidrule(lr){1-4}       
         \multirow{2}{*}{\shortstack{Non-\\Members}}  & \multirow{2}{*}{214} & \multirow{2}{*}{1000}  & \multirow{2}{*}{25.34} \\ 
     & & & \\ 
    \bottomrule
    \end{tabular}
\end{table}

\begin{table}[!t] 
    \centering
    \footnotesize
    \caption{Syntax Convention Counts in our Benchmark}\label{tab:syntax-stats}
    \begin{tabular}{m{4cm}m{2cm}}
    \toprule
    \textbf{Category} & \textbf{Count} \\
    \midrule
    Data Model & 63,594 \\
    Expressions & 30,988 \\
    Single Statements & 321 \\
    Compound Statements & 11,816 \\
    \midrule
    \multicolumn{1}{l}{\textbf{Total Syntax Tokens:}} & 95,257 \\
    \multicolumn{1}{l}{\textbf{Total Tokens:} } & 248,218 \\
    \multicolumn{1}{l}{\textbf{Syntax Tokens Ratio:} } & 38.4\% \\
    \bottomrule
    \end{tabular}
\end{table}

We count the occurrences of the syntax conventions (\ie~conditions and consequences) that existed in our benchmark. Table~\ref{tab:syntax-stats} presents the distribution of these syntax convention categories in our benchmark, which demonstrates the syntactic diversity and complexity of the collected functions. The data model accounts for the largest portion of syntax convention tokens, contributing 63,594 tokens, followed by expressions with 30,988 tokens. In total, the four categories result in 95,257 tokens that would be pruned by \tool for MIA, representing 38.4\% of the overall token set.



\textbf{Ratio Settings.} We evaluated the MIAs under three different member-to-non-member ratios: 1:1, 1:5, and 5:1. For the 1:1 setting, we used 1,000 members and 1,000 non-members. For the 1:5 setting, we randomly sampled 200 members and combined them with 1,000 non-members. Conversely, the 5:1 setting consisted of 1,000 members and a random sample of 200 non-members. These varied distributions were designed to assess the performance and robustness of performing MIAs of code members in LLMs, particularly under imbalanced dataset scenarios.






\section{Experiments}\label{sec:experiments}

We evaluate \tool through comprehensive experiments. We have provided the replicating artifact at \textit{\url{https://anonymous.4open.science/r/SYNPRUNE-FED7/}}.


\subsection{Setup}

\begin{table*}[!t]
\centering
\footnotesize
\caption{AUROC (\%) of different methods under varying member-to-non-member ratios.}
\label{tab:main_results}
\begin{tabular}{m{1cm}m{2cm}>{\centering\arraybackslash}m{2.2cm}>{\centering\arraybackslash}m{2.2cm}>{\centering\arraybackslash}m{2.2cm}>{\centering\arraybackslash}m{2.2cm}}
\toprule
\textbf{Ratio} & \textbf{Method} & \textbf{Pythia 2.8B} & \textbf{GPT-Neo 2.7B}  & \textbf{StableLM-Alpha 3B} & \textbf{GPT-J 6B}\\
\midrule
\multirow{6}{*}{1:1} & \textsc{Loss} & 38.4 & 43.7 & 33.3 & 43.3 \\
& \textsc{ZLib} & 33.2 & 35.5 & 32.0 & 35.7 \\
& \textsc{Min-K} & 38.8 & 41.8 & 34.2 & 41.1 \\
& \textsc{DC-PDD} & 50.1 & 41.2 & 43.7 & 41.3 \\
\cmidrule(lr){2-6}
& \tool & 61.3 & 59.7 & 61.3 & 59.7 \\
\midrule
\multirow{6}{*}{1:5} & \textsc{Loss} & 40.3 & 45.3 & 33.9 & 44.8 \\
& \textsc{ZLib} & 34.0 & 36.2 & 32.5 & 36.3 \\
& \textsc{Min-K} & 40.2 & 43.3 & 34.9 & 42.4 \\
& \textsc{DC-PDD} & 51.8 & 44.0 & 44.6 & 43.2 \\
\cmidrule(lr){2-6}
& \tool & 61.2 & 59.4 & 61.1 & 61.5 \\
\midrule
\multirow{6}{*}{5:1} & \textsc{Loss} & 40.0 & 43.8 & 34.9 & 43.6 \\
& \textsc{ZLib} & 33.1 & 35.0 & 31.9 & 35.2 \\
& \textsc{Min-K} & 39.9 & 42.2 & 35.1 & 41.6 \\
& \textsc{DC-PDD} & 49.4 & 40.9 & 44.1 & 39.6 \\
\cmidrule(lr){2-6}
& \tool & 62.0 & 60.7 & 62.0 & 63.1 \\
\bottomrule
\end{tabular}
\end{table*}


\textbf{Baselines.} We replicate four recent and representative baselines that represent the current state-of-the-art.

\begin{itemize}
    \item \textsc{Loss}~\cite{yeom2018privacy}. \textsc{Loss} uses the overall perplexity (i.e., cross-entropy loss) of a language model as the detection score, based on the standard assumption in membership inference attacks that training data yields lower perplexity than unseen data.
    \item \textsc{ZLib}~\cite{carlini2021extracting}. \textsc{ZLib} applies the ZLib compression algorithm to each sample's tokenized representation, using the compressed length as a detection score.,
    \item \textsc{Min-K\%}~\cite{shi2023detecting}. \textsc{Min-K\%} ranks tokens by likelihood and computes a score using the bottom K (ranging from 10 to 100) based on their aggregated likelihoods. This approach captures the model's uncertainty over less likely tokens, as explored in prior token-level membership inference studies.
    \item \textsc{DC-PDD}~\cite{zhang2024pretraining}. The \textsc{DC-PDD} improves detection performance on pre-training data by calibrating distributional differences, making it suitable for multilingual scenarios.
\end{itemize}




\textbf{Targeted LLMs.} We found evidence that the following four models were trained on the Pile dataset: \textit{EleutherAI/pythia-2.8b}~\cite{pilepythis}, \textit{EleutherAI/gpt-neo-2.7B}~\cite{pilegpt_neo}, \textit{StabilityAI/stablelm-base-alpha-3b}~\cite{pilestable}, and \textit{EleutherAI/gpt-j-6b}~\cite{pilegpt}. Therefore, these four models can be evaluated on member and non-member samples from our constructed benchmark. Specifically, \textit{EleutherAI/gpt-neo-2.7B} is a 2.7B parameter Transformer model designed for text generation and creative tasks. \textit{StabilityAI/stablelm-base-alpha-3b} is a 3B parameter model optimized for code generation and low-resource deployment. \textit{EleutherAI/gpt-j-6b} is a 6B parameter model with enhanced reasoning capabilities. 


\textbf{Metrics.} We use the Area Under the Receiver Operating Characteristic curve (AUROC) as our evaluation metric. AUROC is widely adopted for binary classification tasks and is particularly suitable for membership inference. It captures the trade-off between true positive rate and false positive rate, providing an aggregate judge of the effect of all possible thresholds (i.e., $\epsilon$). It offers a robust measure of a method's ability to distinguish between training and non-training data in language models. We also use the false negative rate (FNR) to measure the percentage of members missed by the detection techniques, and the f1-score to select the most effective threshold in different models.

\textbf{Environment.} Our experiments were conducted on a single NVIDIA RTX 4090D GPU (24GB), supported by a 16-core Intel Xeon Platinum 8474C CPU and 80GB of RAM. The software environment consisted of Ubuntu 20.04, Python 3.8, PyTorch 2.0.0, and CUDA 11.8.

\begin{table}[!t] 
    \centering
    \footnotesize
    \caption{False Negative Rate (\%) of \tool on Members under Short and Long Function Lengths}
    \begin{tabular}{m{1.3cm}>{\centering\arraybackslash}m{1.1cm}>{\centering\arraybackslash}m{1.1cm}>{\centering\arraybackslash}m{1.1cm}>{\centering\arraybackslash}m{1.1cm}}
    \toprule
    \textbf{Length} & \textbf{\shortstack{Pythia\\2.8B}} & \textbf{\shortstack{GPT-Neo\\2.7B}}  & \textbf{\shortstack{StableLM-\\Alpha\\3B}} & \textbf{\shortstack{GPT-J\\6B}}\\
    \midrule
    Short & 2.10 & 19.39 & 0.00 & 0.00 \\
    Long & 75.12 & 75.13 & 66.04 & 87.09 \\
    \bottomrule
    \end{tabular}
    \label{tab:length_results}
\end{table}

\subsection{Results}

Table~\ref{tab:main_results} presents the effectiveness of \tool compared to baseline methods across four models, demonstrating consistent superiority in 1:1, 1:5, and 5:1 ratios. Specifically, \tool achieves average AUROCs of 61.5\%, 59.9\%, 61.4\%, and 61.4\% in Pythia 2.8B, GPT-Neo 2.7B, StableLM-Alpha 3B, and GPT-J 6B, respectively. 
\tool maintains consistent and similar effectiveness across difference model architectures and scales.
Compared to state-of-the-art methods, \tool achieves AUROC improvements ranging from 11.2\% to 17.6\% (1:1 ratio), 9.4\% to 16.7\% (1:5 ratio), and 12.6\% to 19.5\% (5:1 ratio), underscoring its robust performance advantages. Overall, \tool achieves an average AUROC improvement of 15.4\% across four models and three data ratios compared to the state-of-the-art.

\subsection{Robustness under Different Function Lengths}

We categorize the member functions by token lengths into two groups based on statistical distribution. Specifically, we sort all member functions in the benchmark by their token counts in ascending order, then divide them into two groups (500:500) using the median number as the split point: \textit{short} (token count $\leq$ 55) and \textit{long} (token count $>$ 55). This balanced distribution ensures fair comparison under varying function complexities. 

Table~\ref{tab:length_results} shows \tool's effectiveness in terms of false negative rate. The results demonstrate that \tool generates fewer false positives with shorter functions. Notably, \tool achieved perfect member recall when evaluated on both StableLM-Alpha 3B and GPT-J 6B, correctly identifying all members without any misses. Meanwhile, longer functions correlate with increased false negative rates, indicating future improvements of MIA techniques in handling longer functions. 

\subsection{Ablation Study}

We created four ablated versions of \tool by individually removing syntax conventions from each of the following categories: data model (w/o DM), expressions (w/o Expr.), single statements (w/o S.Stmt), and compound statements (w/o C.Stmt). 

We evaluated the four ablated versions on the three ratios, with the results presented in Fig. \ref{tab:ablation}. Each cell shows the difference in AUROC compared to the original \tool. We observe that all ablated versions exhibit an AUROC drop across all three ratios, demonstrating the contribution of the syntax conventions in each category. Our ablation study reveals the impact of syntax convention removal on model performance. Eliminating data model syntax conventions resulted in average AUROC decreases of 3.9\% (1:1), 2.5\% (1:5), and 1.9\% (5:1). Similarly, removing expression syntax conventions led to reductions of 4.0\%, 3.3\%, and 3.5\%, while ablating single statement conventions caused drops of 3.8\%, 3.6\%, and 2.4\% across the respective ratios. The most significant impact came from compound statement syntax removal, with AUROC losses of 5.9\%, 5.6\%, and 4.1\%.


\begin{table}[!t] 
    \centering
    \footnotesize
    \caption{Ablation Study (Each cell indicates the difference in AUROC compared to the original \tool.)}\label{tab:ablation}
    \begin{tabular}{m{0.35cm}m{1.5cm}>{\centering\arraybackslash}m{1cm}>{\centering\arraybackslash}m{1cm}>{\centering\arraybackslash}m{1.1cm}>{\centering\arraybackslash}m{0.95cm}}
    \toprule
    \textbf{Ratio} & \textbf{Method} & \textbf{\shortstack{Pythia\\2.8B}} & \textbf{\shortstack{GPT-Neo\\2.7B}}  & \textbf{\shortstack{StableLM-\\Alpha\\3B}} & \textbf{\shortstack{GPT-J\\6B}}\\
    \midrule
    \multirow{4}{*}{1:1} & w/o DM & $\downarrow$\,5.1 & $\downarrow$\,2.9 & $\downarrow$\,5.8 & $\downarrow$\,1.8 \\
                        &  w/o Expr. & $\downarrow$\,4.5 & $\downarrow$\,3.6 & $\downarrow$\,5.6 & $\downarrow$\,2.4 \\
                        &  w/o S.Stmt & $\downarrow$\,5.0 & $\downarrow$\,2.5 & $\downarrow$\,6.0 & $\downarrow$\,1.6 \\
                        & w/o C.Stmt & $\downarrow$\,6.1 & $\downarrow$\,6.7 & $\downarrow$\,7.1 & $\downarrow$\,3.8 \\
    \midrule
    \multirow{4}{*}{1:5} & w/o DM & $\downarrow$\,4.1 & $\downarrow$\,0.1 & $\downarrow$\,4.7 & $\downarrow$\,1.0 \\
                        &  w/o Expr. & $\downarrow$\,3.1 & $\downarrow$\,2.6 & $\downarrow$\,4.5 & $\downarrow$\,3.1 \\
                        &  w/o S.Stmt & $\downarrow$\,4.4 & $\downarrow$\,2.2 & $\downarrow$\,5.1 & $\downarrow$\,2.6 \\
                        & w/o C.Stmt & $\downarrow$\,4.8 & $\downarrow$\,6.5 & $\downarrow$\,6.2 & $\downarrow$\,4.9 \\
    \midrule
    \multirow{4}{*}{5:1} & w/o DM & $\downarrow$\,2.7 & $\downarrow$\,0.1 & $\downarrow$\,3.6 & $\downarrow$\,1.2 \\
                        &  w/o Expr. & $\downarrow$\,2.6 & $\downarrow$\,4.4 & $\downarrow$\,4.1 & $\downarrow$\,3.0 \\
                        &  w/o S.Stmt & $\downarrow$\,3.2 & $\downarrow$\,0.7 & $\downarrow$\,3.9 & $\downarrow$\,1.8 \\
                        & w/o C.Stmt & $\downarrow$\,4.1 & $\downarrow$\,3.2 & $\downarrow$\,5.2 & $\downarrow$\,3.9 \\
    \bottomrule
    \end{tabular}
\end{table}

\subsection{Threshold Analysis}

\begin{figure}[!t]
\centering
\begin{subfigure}[b]{0.23\textwidth}
\centering
\includegraphics[width=0.99\textwidth]{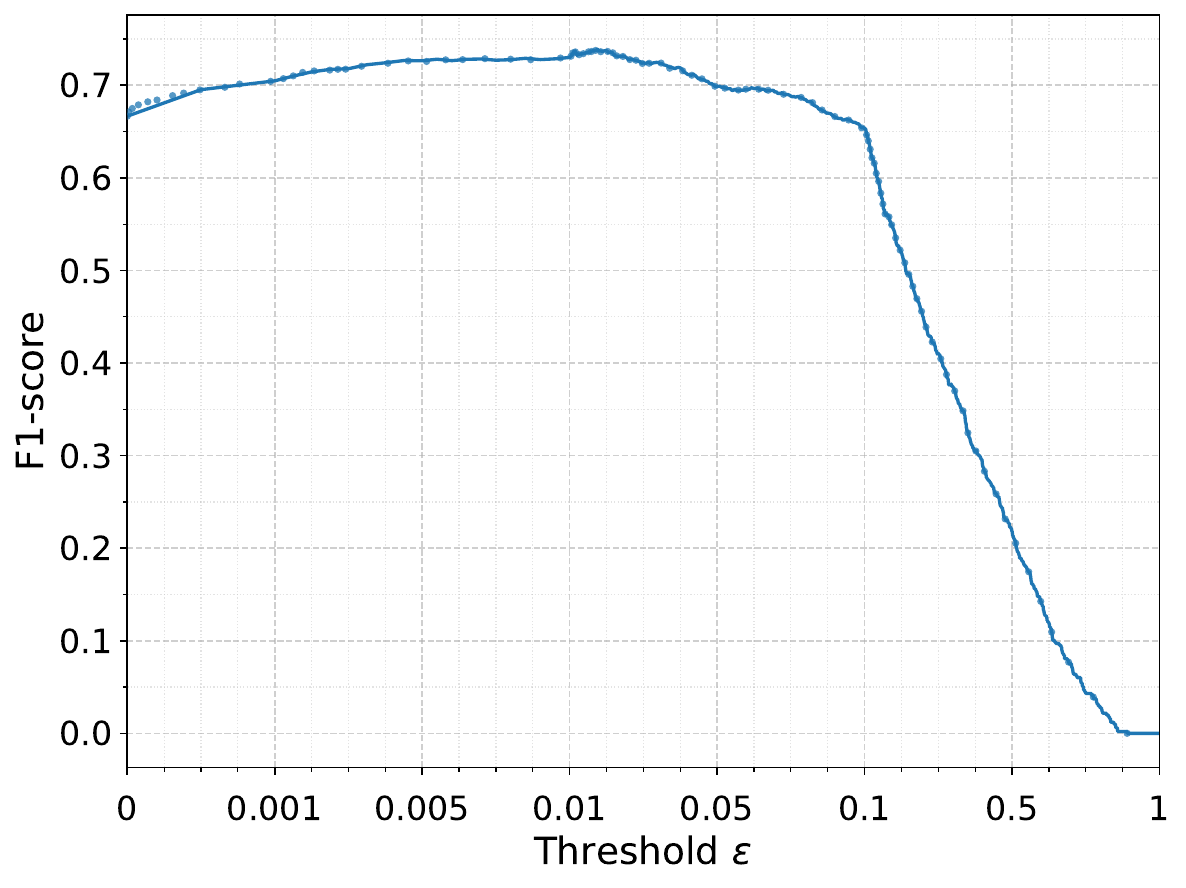}
\caption{Pythia 2.8B}
\label{fig:pythia}
\end{subfigure}
\begin{subfigure}[b]{0.23\textwidth}
\centering
\includegraphics[width=0.99\textwidth]{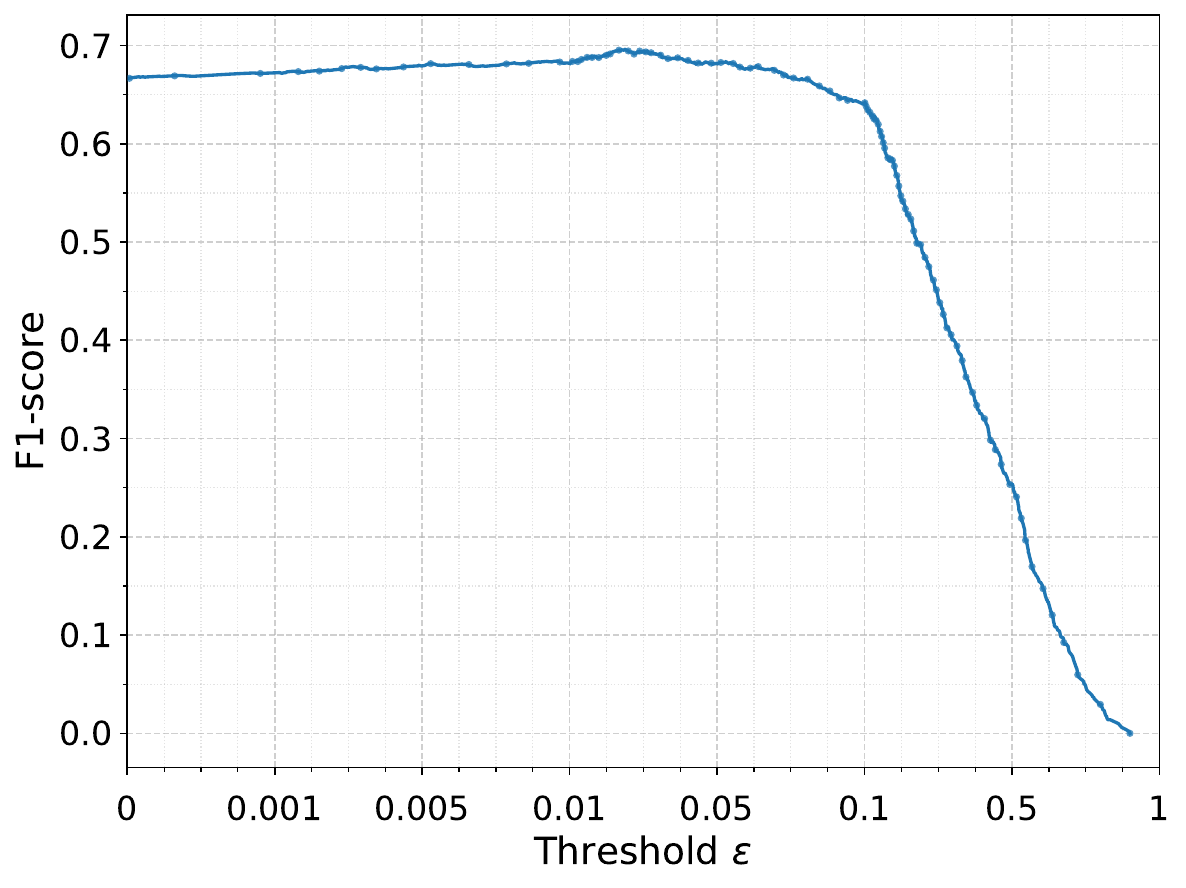}
\caption{GPT-Neo 2.7B}
\label{fig:gpt27}
\end{subfigure}
\begin{subfigure}[b]{0.23\textwidth}
\centering
\includegraphics[width=0.99\textwidth]{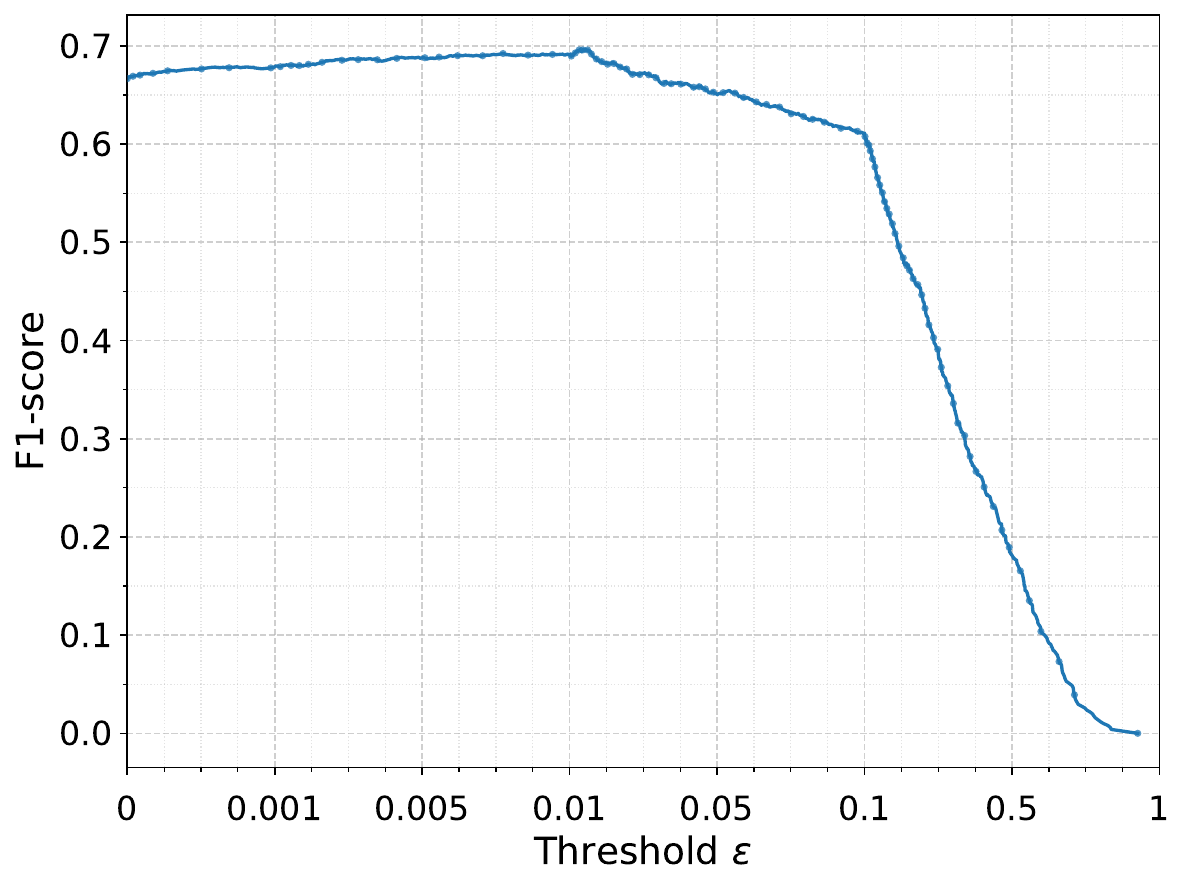}
\caption{StableLM-Alpha 3B}
\label{fig:stable}
\end{subfigure}
\begin{subfigure}[b]{0.23\textwidth}
\centering
\includegraphics[width=0.99\textwidth]{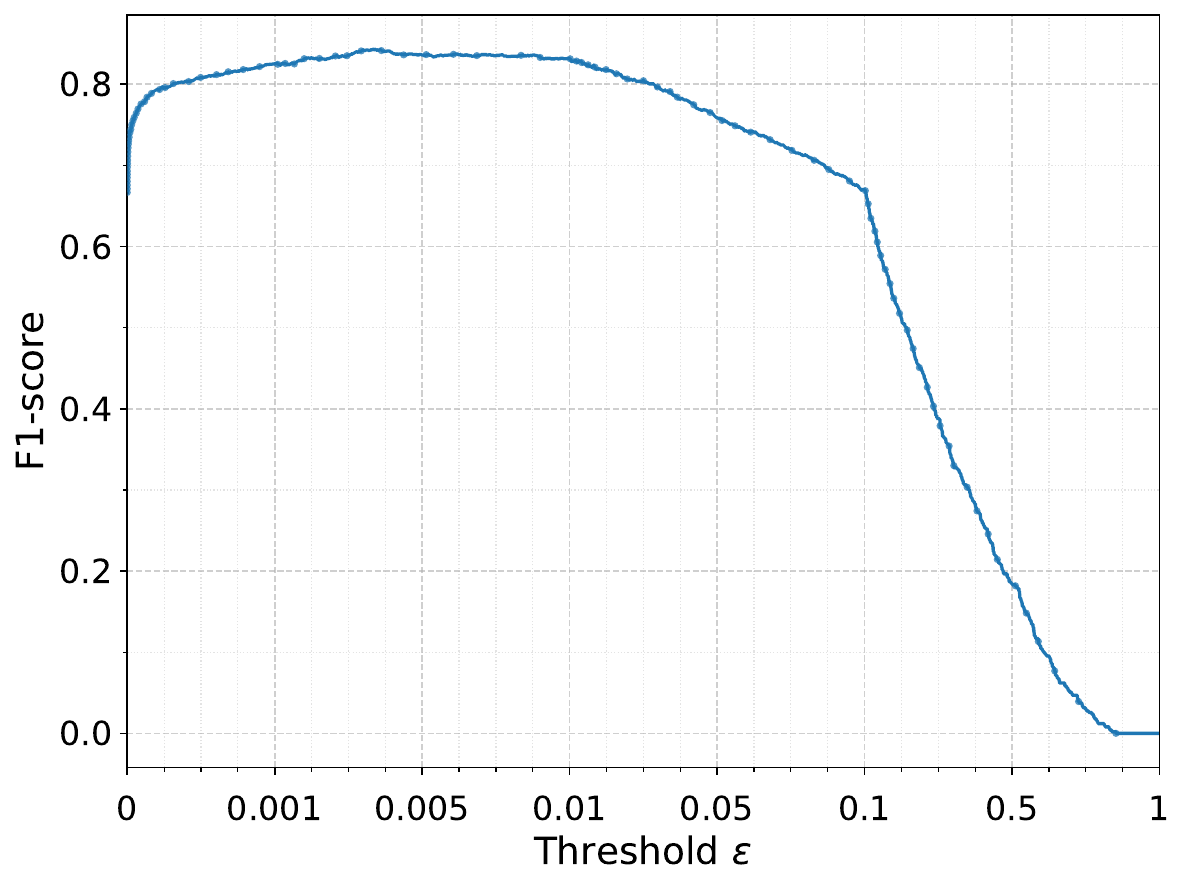}
\caption{GPT-J 6B}
\label{fig:gpt6b}
\end{subfigure}
\caption{F1-Score of \tool Across Varying Thresholds}\label{fig:sensitivity}
\end{figure}

Fig. \ref{fig:sensitivity} presents the curve of f1-score of \tool in the four models across varying thresholds. Note that we applied multiple axis scales to better illustrate the results, as the SPP scores are typically very small. Observing from the graph, we can see that all four models undergo an minor increase of F1-score when $\epsilon$ increases and confront a plunge after $\epsilon$ exceeds around 0.01. We selected the most effective $\epsilon$ for the four models. Specifically, Pythia achieves an F1-score of 0.7390 at $\epsilon = 0.0175$; GPT-Neo reaches 0.6963 at $\epsilon = 0.0250$; StableLM-Alpha records 0.6966 at $\epsilon = 0.0143$; and GPT-J achieves 0.8432 at $\epsilon = 0.0037$.

\subsection{Limitations}

While \tool is currently implemented and evaluated using Python's grammar, its design readily extends to other languages (e.g., Java, C/C++, JavaScript). This generalizability stems from shared language conventions and the universal requirement of strict grammatical correctness across programming languages. We therefore anticipate \tool would achieve similar effectiveness when applied to these other languages. Besides, our evaluation is currently limited to four LLMs due to resource constraints. Nevertheless, the consistent results across all models suggest strong generalizability of our approach to other large language models.


\section{Conclusions}

We introduced \tool, a syntax-pruned MIA technique specifically tailored for detecting pretraining code members from LLMs. By filtering out syntax-constrained tokens from the analysis, \tool focuses on more semantically meaningful signals of memorization. We evaluated \tool on a real-world benchmark of Python functions, carefully constructed to distinguish between verifiable members and non-members. Our results show that \tool consistently outperforms existing MIA baselines, improves robustness across code lengths, and provides practical evidence of memorization in deployed LLMs.

\section*{Ethical Considerations}

This work is motivated by the pressing concern of respecting intellectual property in the development of large language models (LLMs), particularly regarding the use of copyrighted code in pretraining datasets. While membership inference attacks (MIAs) raise privacy and model extraction concerns, we limit our use to research contexts aimed at evaluating potential copyright risks, not for exploitation. We adhere to responsible disclosure norms and aim to foster greater transparency and accountability in LLM development.


\bibliography{src/reference}



\end{document}



\title{Supplementary material: Full List of Syntax Conventions of Python}\label{app:database}

\author{
    Yuanheng Li\textsuperscript{\rm 1}\thanks{Both authors contributed equally to this work.},
    Zhuoyang Chen\textsuperscript{\rm 1}\footnotemark[1],
    Xiaoyun Liu\textsuperscript{\rm 1},
    Yuhao Wang\textsuperscript{\rm 1},\\
    Mingwei Liu\textsuperscript{\rm 2},
    Yang Shi\textsuperscript{\rm 1},
    Kaifeng Huang\textsuperscript{\rm 1}\thanks{Kaifeng Huang is the corresponding author.},
    Shengjie Zhao\textsuperscript{\rm 1}
}

\affiliations{
    \textsuperscript{\rm 1}Tongji University\\
    \textsuperscript{\rm 2}Sun Yat-sen University
}

\maketitle

\begin{table*}[!t]
    \centering
    \footnotesize
    \caption{List of the Collected Syntax Conventions (\texttt{[SP]}, \texttt{[BR]}, and \texttt{[IND]} denote the space, line break, and indentation, resp.)}\label{table:syntax_full}
    \begin{tabular}{m{2.8cm}m{2cm}m{6.5cm}m{5cm}}
        \toprule
        \textbf{Category} &\textbf{Syntax Node}  & \textbf{Conditional Token} & \textbf{Consequent Token} \\
        \midrule
        
        \multirow{7}{*}{Data Model}  & Lists, Slice & `[' & `]' \\
        \cmidrule{2-4}
            & Dict, Set &  `\{' & `\}' \\
            \cmidrule{2-4}
            & String, Bytes  & ``',`"' &   `'',`"'  \\
            \cmidrule{2-4}
            & Object & \textit{identifier} & `.' \\
            \cmidrule{2-4}
            & Tuple & `(' & `)' \\\hline
        \multirow{6}{*}{Expressions} & call & \textit{identifier}( & `)' \\
        \cmidrule{2-4}
            & lambda & \texttt{lambda} \textit{params} & `:' \\
        \cmidrule{2-4}
            & conditional & \textit{expr} \texttt{if} \textit{cond} & \texttt{else} \\
        \cmidrule{2-4}
            & \multirow{2}{*}{comprehension} & \textit{expr} \texttt{for} \textit{target} & \texttt{in}  \\
        \cmidrule{3-4}
            & & \textit{expr} \texttt{for} \textit{target} \texttt{in} \textit{iterable} & `]', `\}', `)'  \\
        \cmidrule{2-4}
            & \shortstack[l]{chained\\comparison} & \textit{expr comp\_op expr} & \textit{comp\_op} \\\hline
        \multirow{4}{*}{Single Statements} 
        & \multirow{2}{*}{\texttt{import}} & \texttt{import} \textit{module}  & \texttt{as} \\
        \cmidrule{3-4}
        & & \texttt{from} \textit{module} & \texttt{import} \\
        \hline
        \multirow{31}{*}{Compound Statements} 
        & \multirow{3}{*}{\texttt{if}} & \texttt{if} \textit{assignment\_expression} & `:'\\
        \cmidrule{3-4}
        & & \texttt{if} \textit{assignment\_expression}\text{:} & \texttt{[SP]}, \texttt{[BR]}, \texttt{[IND]} \\
        \cmidrule{3-4}
        & & \texttt{if} \textit{assignment\_expression}\text{:} & \texttt{elif}, \texttt{else}\\
        \cmidrule{2-4}
        & \multirow{3}{*}{\texttt{for}}   & \texttt{for} \textit{target\_list}  & \texttt{in}\\
        \cmidrule{3-4}
        & & \texttt{for} \textit{target\_list} \texttt{in} \textit{starred\_list}  & `:' \\
        \cmidrule{3-4}
        & & \texttt{for} \textit{target\_list} \texttt{in} \textit{starred\_list}\text{:} & \texttt{[SP]}, \texttt{[BR]}, \texttt{[IND]}  \\
        \cmidrule{2-4}
        & \multirow{7}{*}{\texttt{try}} &  \texttt{try} &  `:' \\
        \cmidrule{3-4}
        & & \texttt{try:} & \texttt{[SP]},  \texttt{[BR]}, \texttt{[IND]} \\
        \cmidrule{3-4}
        & & \texttt{try:} \textit{suite} & \texttt{except}, \texttt{except*} \\
        \cmidrule{3-4}
        & & \texttt{try:} \textit{suite} \texttt{except} \textit{expression}  &  \texttt{as} \\
        \cmidrule{3-4}
        & & \texttt{try:} \textit{suite} \texttt{except} \textit{expression} \texttt{as} \textit{name} & `:' \\
         \cmidrule{3-4}
         & & \texttt{try:} \textit{suite} \texttt{except} \textit{expression} \texttt{as} \textit{name:} & \texttt{[SP]},  \texttt{[BR]}, \texttt{[IND]}\\
         \cmidrule{3-4}

        & & \texttt{try:} \textit{suite} \texttt{except} \textit{expr} \texttt{as} \textit{identifier:} \textit{suite} & \texttt{finally}, \texttt{else} \\
        \cmidrule{2-4}
        & \multirow{3}{*}{\texttt{with}} & \texttt{with} \textit{with\_item} & \texttt{as} \\
        \cmidrule{3-4}
        & & \texttt{with} \textit{with\_item} \texttt{as} \textit{target} & `:' \\
        \cmidrule{3-4}

        & & \texttt{with} \textit{with\_item}\text{:} & \texttt{[SP]}, \texttt{[BR]}, \texttt{[IND]} \\
        \cmidrule{3-4}

        & & \texttt{with} \textit{with\_item} \texttt{as} \textit{target}\text{:} & \texttt{[SP]}, \texttt{[BR]}, \texttt{[IND]} \\
        \cmidrule{2-4}

        & \multirow{3}{*}{\texttt{class}}  & \texttt{class} \textit{classname} & `:' \\
        \cmidrule{3-4}
        & & \texttt{class} \textit{classname}\text{:} & \texttt{[SP]}, \texttt{[BR]}, \texttt{[IND]}  \\
        \cmidrule{3-4}
        & & \texttt{class} \textit{classname}(\textit{identifier} &  `,', `)'  \\
        \cmidrule{3-4}
         & & \texttt{class} \textit{classname}(\textit{identifier}) &   `:'  \\
        \cmidrule{2-4}
        & \multirow{5}{*}{function} & \texttt{def} \textit{funcname}( & \texttt{self}, `)' \\
        \cmidrule{3-4}
        & & \texttt{def} \textit{funcname}(\textit{identifier} & `,', `)' \\
        \cmidrule{3-4}
        & & \texttt{def} \textit{funcname}(\textit{parameters}) &  `->' \\
        \cmidrule{3-4}
        & & \texttt{def} \textit{funcname}(\textit{parameters}) [\texttt{->} \textit{type}] & `:' \\
        \cmidrule{3-4}
        & & \texttt{def} \textit{funcname}(\textit{parameters})\text{:} & \texttt{[SP]}, \texttt{[BR]}, \texttt{[IND]}  \\
        \cmidrule{2-4}
        & \multirow{3}{*}{\texttt{while}} & \texttt{while} \textit{condition} & `:'\\
        \cmidrule{3-4}
        & & \texttt{while} \textit{condition}\text{:} & \texttt{[SP]}, \texttt{[BR]}, \texttt{[IND]}  \\
        \cmidrule{3-4}
        & & \texttt{while} \textit{condition}\text{:} \textit{suite}& \texttt{else}\\
        \cmidrule{2-4}
        & \multirow{2}{*}{\texttt{match}} & \texttt{match} \textit{subject} & `:' \\
        \cmidrule{3-4}
        & & \texttt{match} \textit{subject}\text{:} &\texttt{[SP]}, \texttt{[BR]}, \texttt{[IND]}, \texttt{case}  \\
        
        \bottomrule
    \end{tabular}
\end{table*}

